\documentclass[showpacs,amsmath,amssymb,twocolumn]{revtex4}
\usepackage{pstricks,graphicx,amsmath,bbm,mathrsfs,amssymb,psfrag,pifont}
\usepackage{color}
\begin{document}
\title{Characterizing the non-classicality of mesoscopic optical twin-beam states}

\author{Alessia~Allevi}
\email{alessia.allevi@uninsubria.it}
\affiliation{Dipartimento di Scienza e Alta Tecnologia, Universit\`a degli Studi dell'Insubria, I-22100 Como, Italy}
\affiliation{CNISM UdR Como, I-22100 Como, Italy}

\author{Marco~Lamperti}
\affiliation{Dipartimento di Scienza e Alta Tecnologia, Universit\`a degli Studi dell'Insubria, I-22100 Como, Italy}

\author{Maria~Bondani}
\affiliation{Istituto di Fotonica e Nanotecnologie, CNR-IFN, I-22100 Como, Italy}
\affiliation{CNISM UdR Como, I-22100 Como, Italy}

\author{Jan Pe\v{r}ina,~Jr.}
\address{Institute of Physics of Academy of Sciences of the Czech
Republic, Joint Laboratory of Optics of Palack\'{y} University,
17. listopadu 12, 772 07 Olomouc, Czech Republic}

\author{Ond\v{r}ej Haderka}
\address{RCPTM, Joint Laboratory of Optics of Palack\'{y} University
and Institute of Physics of Academy of Sciences of the Czech
Republic, Faculty of Science, Palack\'{y} University, 17.
listopadu 12, 77146 Olomouc, Czech Republic}

\author{Radek Machulka}
\address{RCPTM, Joint Laboratory of Optics of Palack\'{y} University
and Institute of Physics of Academy of Sciences of the Czech
Republic, Faculty of Science, Palack\'{y} University, 17.
listopadu 12, 77146 Olomouc, Czech Republic}

\author{V\'{a}clav Mich\'{a}lek}
\address{Institute of Physics of Academy of Sciences of the Czech
Republic, Joint Laboratory of Optics of Palack\'{y} University,
17. listopadu 12, 772 07 Olomouc, Czech Republic}

%%%%%%%%%%%%%%
\begin{abstract}
We present a robust tool to analyze nonclassical properties of
multimode twin-beam states in the mesoscopic photon-number
domain. The measurements are performed by direct detection. The
analysis exploits three different non-classicality criteria for
detected photons exhibiting complementary behavior in the explored
intensity regime. Joint signal-idler photon-number distributions
and quasi-distributions of integrated intensities are determined
and compared with the corresponding distributions of detected
photons. Experimental conditions optimal for nonclassical
properties of twin-beam states are identified.
\end{abstract}
%%%%%%%%%%%%%%%%%%%%%%%%%%%%%%%%%%%%%%%%%%%%%%%%%%%%%%%
\maketitle
%%%%%%%%%%%%%%%%%%%%%%%%%%%%%%%%%%%%%%%%%%%%%%%%%%%%%%%%%%%%%
\section{Introduction}
\label{s:intro} Quantum nature of an optical state is mandatory
for exploiting the state in many useful applications including
those in quantum information and metrology
\cite{kim02,braunstein05,walmsley05,ralph06,kok07}. By definition,
a state is nonclassical whenever it cannot be written as a
positive superposition of coherent states. Using the
Glauber-Sudarshan representation of a statistical operator
\cite{glauber63,sudarshan63}, nonclassical states are described by
negative or even singular probability $P$-functions
(quasi-distributions). However, as $P$-functions introduced in
this representation cannot be directly observed, also other
non-classicality criteria based on measurable quantities have
been derived \cite{klyshko96,richter02,zavatta07,nori10,brida11}.
For instance, the negativity of the Wigner function of a state
available experimentally is commonly used as a non-classicality
indicator \cite{lvovsky01,zavatta04,ourjoumtsev06}. Unfortunately,
this function is defined only for single-mode states and so it
cannot be used to describe the usual spectrally and spatially
multimode fields \cite{lvovsky07,mauerer09}. Moreover, the retrieval of Wigner function,
typically obtained through optical homodyne tomography, is in
general challenging as it requires optimal spatio-temporal
matching between the state under investigation and a local
oscillator \cite{wasi06,zavatta06,polycarpou12}.\\
An alternative approach to investigate the quantum properties of a
state is provided by the direct detection of the number of photons in
the state. Direct detection offers the possibility to reconstruct
the photon-number distribution and evaluate possible correlations
between the components of a bipartite state \cite{ivanova06,avenhaus10,christ11}. The non-classicality of a
photon-number distribution can be indicated by the values of its Fano
factor $F=\sigma^2(n)/ \langle n \rangle$ ($\sigma^2$ and
$\langle\rangle$ stand for variance and mean value, respectively):
$F<1$ means nonclassical sub-Poissonian statistics \cite{mandel,PerinaJr2013b}.
On the other hand, when a bipartite state exhibits photon-number
correlations, a noise reduction factor $R=\sigma^2(n_1 - n_2)/
\langle n_1 +n_2 \rangle$ ($n_1$ and $n_2$ are the signal and
idler photon numbers) having values
lower than 1 indicates non-classicality \cite{smithey92,rarity92,wenger04,Perina2007,Perina2009}.\\
As one has no direct access to photons, it is of paramount
importance to define non-classicality criteria in terms of
detected photons. In fact, the introduction and exploitation of
non-classicality conditions for measurable quantities give the
possibility to avoid the use of photon-number reconstruction
methods that are in general complex.
In this paper, we experimentally investigate optical multimode
twin-beam (TWB) states containing sizeable numbers of photon
pairs. We report on the characterization of their quantumness by
means of a direct detection scheme involving two photon-counting
detectors that are able to operate in the mesoscopic photon-number domain, in which
more than one pair of photons is produced at each laser shot. 
In particular, we compare three different
non-classicality criteria based on detected photon-number
correlations and discuss the conditions suitable for their
application. Moreover, we compare these criteria with the genuine
definition of non-classicality using both the measured joint
signal-idler detected-photon distributions and
reconstructed joint signal-idler photon-number distributions and the corresponding quasi-distributions of integrated intensities \cite{PerinaJr2013}. \\
Even if the overall detection efficiency of our apparatus is
relatively low, we demonstrate that quantities determined for
detected photons are sufficient to reveal the quantum features of the generated TWB states. The presented comprehensive approach can
thus be considered as a robust tool for discriminating
nonclassical TWB states in different experimental regimes.

The paper is organized as follows. Experimental setup is described
in Sec.~II. Nonclassical characteristics of twin-beams derived for
detected photons are analyzed in Sec.~III. Sec.~IV is devoted to the
reconstruction of joint signal-idler photon-number distributions, the determination of
quasi-distributions of integrated intensities and their
nonclassical features. Conclusions are drawn in Sec.~V.

\section{Experimental implementation of multimode TWB states}
\label{s:TWB}
\begin{figure}[h]
\includegraphics[width=7.5cm]{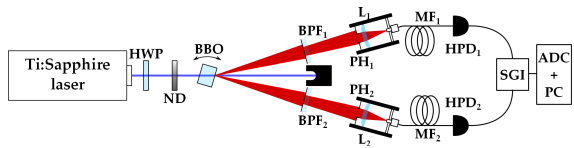}
\caption{Color online. Scheme of the experimental setup. HWP: half-wave plate;
ND: neutral density filter; BBO: nonlinear crystal; BPF$_j$:
bandpass filter; PH$_j$: iris with variable aperture; L$_j$: lens;
MF$_j$: multimode fiber; HPD$_j$: hybrid
photodetector.}\label{f:setup}
\end{figure}

According to the experimental setup shown in Fig.~\ref{f:setup},
mesoscopic TWB states were obtained in spontaneous parametric
down-conversion (SPDC) in a nonlinear crystal with $ \chi^{(2)} $
susceptibility. In particular, we sent the third harmonics (at 266
nm) of a cavity-dumped Kerr-lens mode-locked Ti:Sapphire laser
(Mira 900, Coherent Inc. and PulseSwitch, A.P.E.) to a type I
$\beta$-BaB$_2$O$_4$ crystal (BBO hereafter, 8x8x5~mm$^3$, cut
angle $\vartheta_c=48~\deg$) tuned for slightly non-collinear
interaction geometry. 100-fs long pump-beam pulses were delivered
at frequency 11~kHz.
%%%%%%%%%%%%%%%%
%%%%%%%%%%%%%%%%

\noindent The TWB states generated by the apparatus are
intrinsically multimode, both in spatial and spectral domains. By
assuming that the output energy is equally distributed among $\mu$
modes in each beam, the overall multimode state can be written as
a tensor product of $\mu$ identical single-mode twin-beam states
\cite{allevi10a,allevi12,Perina1991,Perina2005}, \emph{i.e.},
\begin{equation}
|\psi_{\mu}\rangle= \sum_{n=0}^\infty \sqrt{p^\mu_n} |n^{\otimes}\rangle\otimes|n^{\otimes}\rangle, \label{eq:psiTWB}
\end{equation}
where $|n^{\otimes}\rangle =
\delta(n-\sum_{h=1}^{\mu}n_h)\,\otimes_{k=1}^{\mu}|n_k \rangle$
represents an $n$-photon state coming from $\mu$ equally-populated
modes that impinge on the detector and
\begin{equation}
p^\mu_n= \frac{(n +\mu-1)!}{{n!(\mu - 1)! (N / \mu+1)^{\mu} (\mu/N+1)^{n}}} \label{eq:multithermal}
\end{equation}
is a multimode thermal photon-number distribution having $N=
\langle n \rangle $ mean photons \cite{paleari04}. The TWB state
in Eq.~(\ref{eq:psiTWB}) exhibits photon-number correlations that
are provided by pairwise character of SPDC. To investigate the
nature of such correlations and describe their properties, we
collected two frequency-degenerate (at 532 nm) parties of the TWB
state using two symmetric cage systems. The light in each arm was
spectrally filtered by a bandpass filter at high transmissivity,
spatially selected by an iris with variable aperture, focused by a
lens ($f=30$~mm) into a multimode fiber (600-$\mu$m-core
diameter) and delivered to the photodetector. In particular, we used a pair
of hybrid photodetectors (HPD, mod.
R10467U-40, Hamamatsu, Japan). These detectors
are composed by a photocathode, whose quantum efficiency is about $50\%$ in the
investigated spectral region \cite{bondani09a,bondani09b}, followed by an avalanche diode operated below breakdown threshold. 
The internal amplification has a gain profile narrow enough to allow photon-number resolution.
The output of each HPD was amplified (preamplifier A250 plus amplifier
A275, Amptek), synchronously integrated (SGI, SR250, Stanford) and
digitized (ADC, PCI-6251, National Instruments). To perform a
systematic characterization of the generated TWB states, each
experimental run was repeated 200,000 times for fixed choices of
pump mean power and iris sizes.

\section{Nonclassical characteristics of detected photons}
\label{s:detected}  
\begin{figure}[h]
\begin{center}
\includegraphics[width=7.5cm]{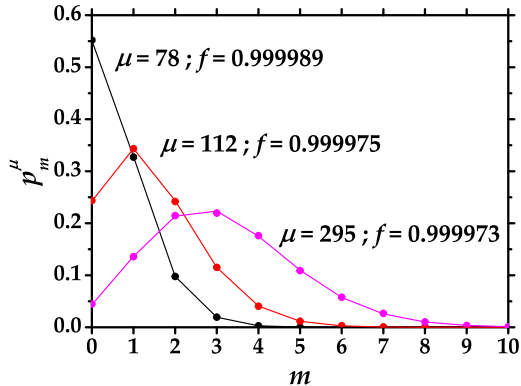}
\end{center}
\vspace{-0.8cm} \caption{Color online. Experimental detected-photon-number
distribution in the signal arm for three different values of pump
mean power (black dots: 49.2~$\mu$W, $\langle m \rangle = 0.60$ and $\mu =78$; red dots:
118.1~$\mu$W, $\langle m \rangle = 1.42$ and $\mu = 112$; magenta dots: 258.3~$\mu$W, $\langle m \rangle
= 3.14$ and $\mu = 295$) for the fixed value of iris sizes (46~mm$^2$), lines:
theoretical expectations. Fidelities
in the figure are calculated as
$f=\sum_m{\sqrt{p^\mu_{m,exp}p^\mu_{m,th}}}$, where the subscript $ exp $
($ th $) denotes experimental (theoretical) distributions. Error
bars are smaller than the symbol sizes.} \label{statistics}
\end{figure}

\begin{figure}[h]
\begin{center}
\includegraphics[width=7.5cm]{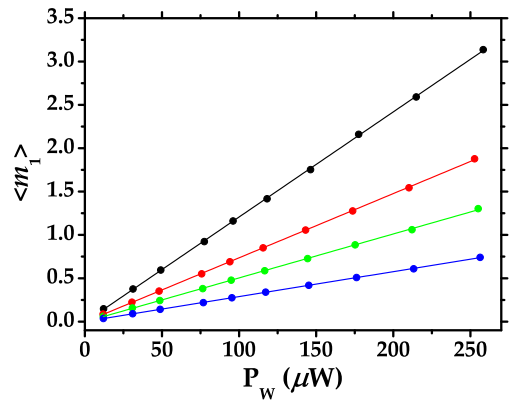}
\end{center}
\vspace{-0.8cm} \caption{Color online. Mean number of detected photons in the signal
arm as a function of pump mean power ${\rm P_w}$ for different values of iris
sizes (from top to bottom: black: 45.92 mm$^2$, red: 20.58 mm$^2$, green: 10.63
mm$^2$, blue: 5.67 mm$^2$). Dots: experimental data; lines: linear
fitting curves.} \label{linear}
\end{figure}

By exploiting the self-consistent analysis
method extensively described in \cite{bondani09a,andreoni09}, we
processed the output of each detection chain, obtained
detected-photon-number distributions and evaluated shot-by-shot
photon-number correlations. In accordance with
Eq.~(\ref{eq:multithermal}) and by taking into account invariance
of the functional form of statistics under Bernoullian detection
\cite{bondani09c}, the detected photon-number distributions are
described by multimode thermal distributions, in which the number
of modes can be determined as $\mu=\langle m \rangle^2/
(\sigma^2(m)- \langle m \rangle)$ \cite{Perina1991,bondani09b},
where $m=\eta n$ stands for the number of detected photons, $ n $
denotes the number of photons and $\eta$ is the quantum detection
efficiency. In Fig.~\ref{statistics} we plot the experimental
detected-photon-number distributions in the signal arm for three
different values of the pump-beam power keeping fixed the value of
iris size (dots). Lines are the expected theoretical curves obtained from
Eq.~(\ref{eq:multithermal}) by replacing $N$ by the measured mean
number of photons. The mean detected-photon numbers presented in
Fig.~\ref{statistics} demonstrate the capability of the detection
apparatus to capture TWB states in different intensity regimes.
Nevertheless, it is worth noting that the SPDC gain is linear in
the whole investigated photon-number domain. This is evident in
Fig.~\ref{linear}, where we show the mean values of photons
detected in the signal arm as functions of the pump mean power for
different values of iris sizes.
%%%%%%%%%
%%%%%%%%%
%%%%%%%%%
%%%%%%%%%

\begin{figure}[h]
\begin{center}
$(a)$\\
\includegraphics[width=7.5cm]{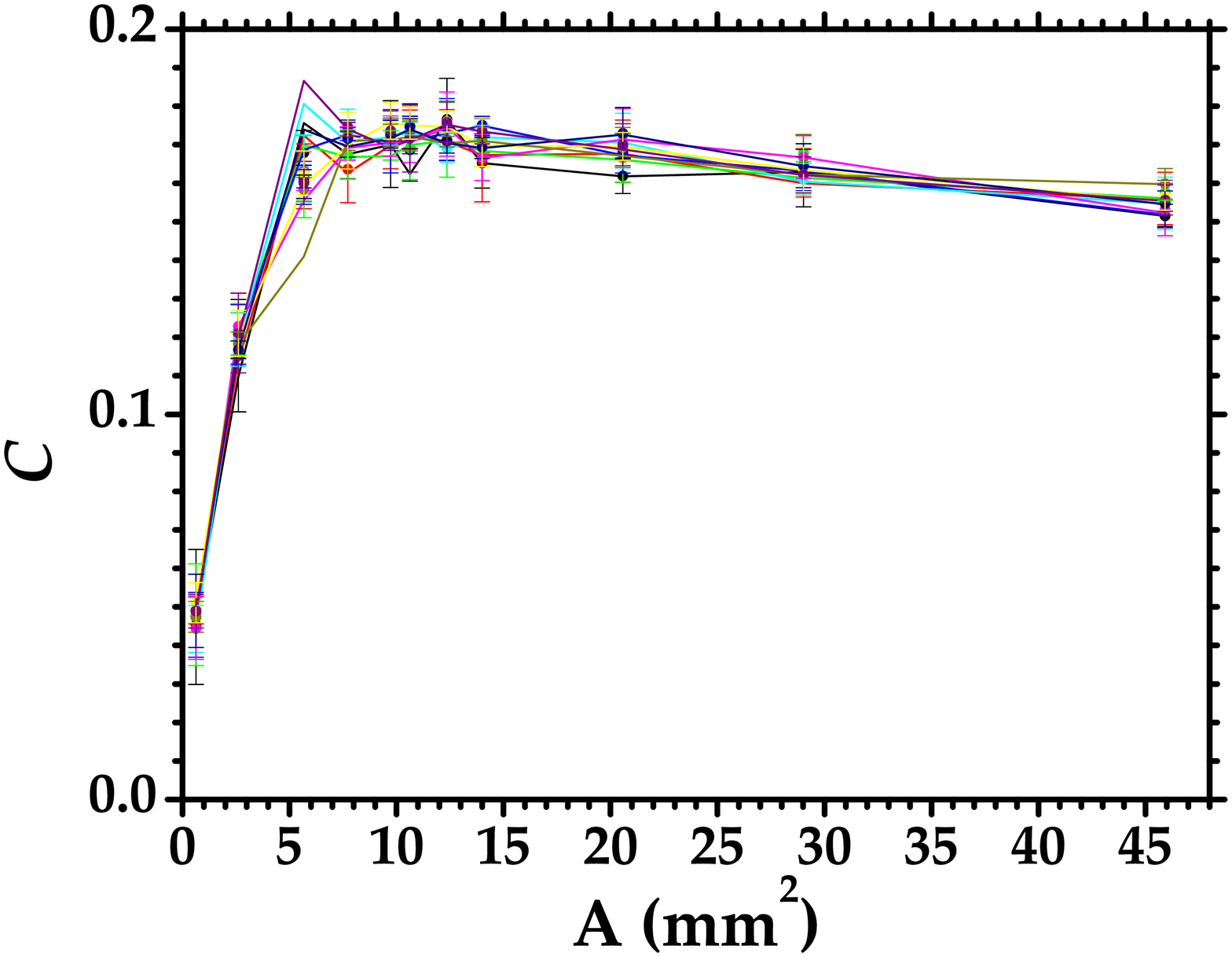}\\
\vspace{0.3cm}
$(b)$\\
\includegraphics[width=7.5cm]{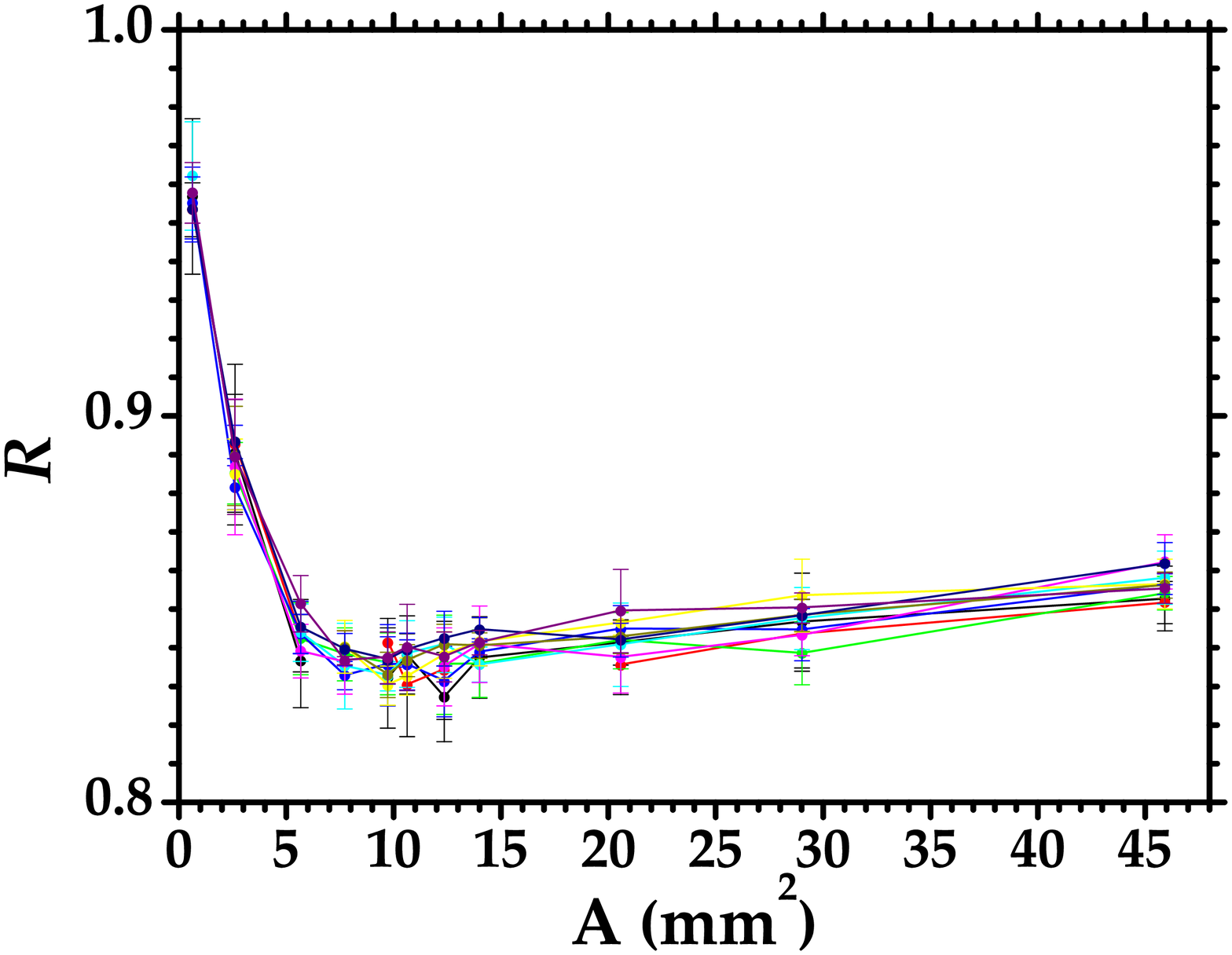}
\end{center}
\vspace{-0.6cm} \caption{Color online. $(a)$, intensity correlation coefficient $C$ and $(b)$, noise reduction factor $ R $ as functions of
iris sizes ${\rm A}$ for different values (different colors) of pump mean
power. Dots: experimental data; lines: theoretical expectations. The lines are used to better guide the eye.}
\label{NRF}
\end{figure}

\begin{figure}[h]
\begin{center}
\includegraphics[width=7.5cm]{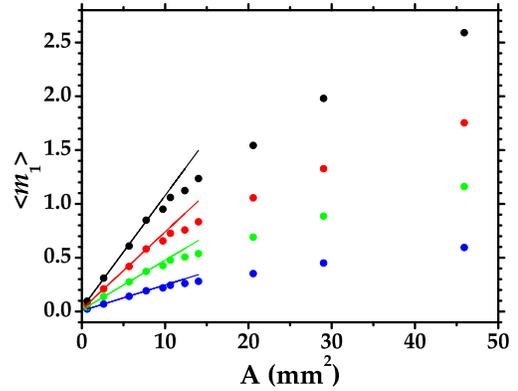}
\end{center}
\vspace{-0.8cm} \caption{Color online. Mean number of detected photons $ \langle
m_1\rangle $ in the signal arm as a function of iris sizes ${\rm A}$ for
different values of pump mean power (from top to bottom: black: 215 $\mu$W, red: 145
$\mu$W, green: 95 $\mu$W, blue: 50 $\mu$W). Dots: experimental
data; lines: linear fitting curves.} \label{meanvalue}
\end{figure}

\noindent The observed detected-photon-number correlations were
quantified by means of the correlation coefficient
\begin{equation}
C =\frac{\langle m_1 m_2 \rangle - \langle m_1\rangle \langle m_2 \rangle}{\sqrt{\sigma^2(m_1)\sigma^2(m_2)}} .  \label{eq:correl}
\end{equation}
that is plotted in Fig.~\ref{NRF}$(a)$ as a function of the value of iris sizes
\cite{bondani07}. However, as already demonstrated in
\cite{agliati05}, the existence of correlations is not sufficient
to discriminate between quantum and classical states. For example,
bipartite states obtained by dividing classical super-Poissonian states at a beam splitter also display photon-number correlations \cite{allevi10b,allevi11}.\\
The noise reduction factor $R$ mentioned above is an explicit
marker of non-classicality originating in photon-number
correlations. For detected photons it is determined along the
formula
\begin{equation}
R=\frac{\sigma^2(m_1- m_2)}{\langle m_1+m_2 \rangle} . \label{eq:NRF}
\end{equation}
It has been shown \cite{degiovanni07} that whenever the value of $R$ lies
in between $1- \eta$ and 1 \cite{bondani07}, the detected state is
nonclassical. In this case, we have sub-shot-noise correlations
since the fluctuations in the detected photon-number correlations
are below the shot-noise level \cite{jedrkiewicz04,brida09,ruo10}.
The behavior of $R$ as a function of the value of iris sizes is
quantified in Fig.~\ref{NRF}$(b)$, in which the nonclassical
character of all obtained data is confirmed \cite{lamperti}. To produce the
theoretical values shown in Fig.~\ref{NRF}, we inserted in
Eqs.~(\ref{eq:correl}) and (\ref{eq:NRF}) the experimental values
of $\eta$, $\langle m_1 \rangle$, $\langle m_2 \rangle$ and $\mu$
obtained in a self-consistent way \cite{allevi12} for each
considered value of the iris sizes. This results in the irregular
behavior of the curve connecting the obtained points in the graphs
in Fig.~\ref{NRF}. Comparison of the curves in Figs.~\ref{NRF}$(a)$
and \ref{NRF}$(b)$ reveals complementary behavior of values of the
correlation coefficient $C$ and noise reduction factor $R$.
Moreover, it follows from the curves in Fig.~\ref{NRF}$(b)$ that the
noise reduction factor $R$ attains its minimum for a certain value
of iris sizes.
%%%%%%%%%
%%%%%%%%%
This occurs when the irises are $\sim 3$-mm wide and select the
largest possible portions of the twin-beam cones
\cite{agafonov10}. This explanation is confirmed by the behavior
of mean detected-photon numbers $\langle m_1 \rangle$ in the
signal arm depending on the iris sizes. As shown in
Fig.~\ref{meanvalue} the mean detected-photon numbers $\langle m_1
\rangle$ stop increasing linearly with the iris size at the same value.
Also the maximum extension of emission cones
beyond the filters was reached in the horizontal plane at this
value. Further increase in mean detected-photon numbers $\langle
m_1 \rangle$ is caused only by additional contributions in the
vertical plane.
%%%%%%%%%
%%%%%%%%%
The values of $C$ and $R$ plotted in Fig.~\ref{NRF} may be
divided into three groups depending on different values of iris
sizes. For small values of the iris sizes, $C$ and $R$ get smaller
and higher values, respectively, as only a small portion of the
twin beam is collected. For moderate values of the iris sizes,
$C$ and $R$ reach their highest and smallest values,
respectively, due to optimum collection conditions. For large
values of the iris sizes, smaller values of $C$ together with
greater values of $R$ are observed because the irises exceed
the width of the cone. \\
We discuss advantages and limitations of the noise reduction
factor $R$ as nonclassicality quantifier in comparison with other
two quantities. In particular, we consider a ratio $ S $ derived
from the Schwarz inequality \cite{vogel06} for detected photons:
\begin{equation}
 S = \frac{\langle m_1 m_2\rangle}{\sqrt{\langle m_1^2\rangle \langle
m_2^2\rangle}}.\label{eq:schw}
\end{equation}
If $ S > 1 $ the state is nonclassical. The second analyzed
quantity is determined from a more recent criterion based on
higher-order detected-photon-number correlations \cite{allevi12}:
\begin{equation}
H = \langle m_1\rangle \langle m_2\rangle\, \frac{g^{22} -
[g^{13}]_{\rm s}}{g^{11}}+ \sqrt{\langle m_1\rangle \langle
m_2\rangle}\,\, \frac{[g^{12}]_{\rm s}}{g^{11}},
 \label{eq:ineq}
\end{equation}
where $ g_m^{jk} = \langle m_1^j m_2^k\rangle\left( \langle
m_1\rangle^j\langle m_2\rangle^k \right)^{-1}$ is the
$(j+k)$th-order correlation function and $[g^{jk}]_{\rm
s}=(g^{jk}+g^{kj})/2$ represents its symmetrized version. If $ H >
1 $ the state is nonclassical.
%%%%%%%%%
\begin{figure}[h!]
\begin{center}
\includegraphics[width=7.5cm]{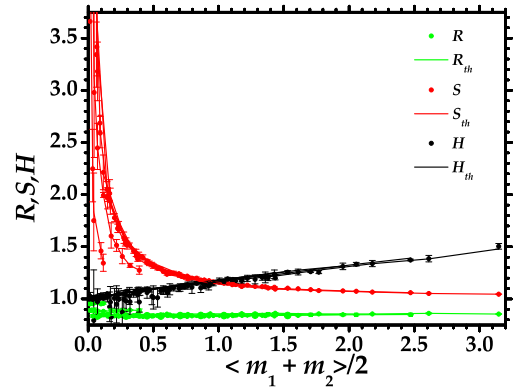}
\end{center}
\vspace{-0.8cm} \caption{Color online. Noise reduction factor $ R $, green color (light gray),
Schwarz-inequality factor $ S $, red color (gray), and higher-order-moments factor $
H $, black color (black), as functions of mean number of photons detected in the two
arms. Dots: experimental data; lines: theoretical expectations,
indicated by subscript $ th $ in the legend.} \label{inequality}
\end{figure}
%%%%%%%%%
In Fig.~\ref{inequality}, we show the results obtained by applying
the above non-classicality criteria to the experimental data. The
three quantities are plotted as functions of the mean number of
photons detected in one of the two arms: good quality of our data
is confirmed by the fact that all criteria are satisfied
simultaneously. For each criterion the data are distributed into
three groups differing in iris sizes, as already mentioned in the
description of Fig.~\ref{NRF}. It is also interesting to note that
all the experimental points (except a very few of them) obtained for different values of pump
mean powers and iris sizes are in good agreement with the
corresponding theoretical predictions calculated for the actual
values of experimental parameters. In particular, the theoretical
curve of noise reduction factor $R$ was drawn along the formula
\begin{equation}
R=1-2\eta\frac{\sqrt{\langle m_1\rangle \langle m_2 \rangle}}{\langle m_1\rangle+
\langle m_2\rangle}+\frac{(\langle m_1\rangle \langle m_2 \rangle)^2}{
\mu(\langle m_1\rangle +\langle m_2\rangle)},\label{eq:Rteo}
\end{equation}
that represents a generalization of the expression derived in
\cite{agliati05} to the multimode case. In Eq.~(\ref{eq:Rteo}),
$\mu$ gives the average of the signal and idler mode numbers,
$\langle m_1\rangle$ and $\langle m_2\rangle$ are the experimental
mean signal and idler detected-photon numbers and a common quantum
detection efficiency $\eta$ was determined from the formula
$R=\sigma^2(m_1-m_2)/(\langle m_1\rangle+ \langle m_2\rangle) =
1-\eta$ valid for an ideal twin beam \cite{allevi12}. As the
curves in Fig.~\ref{inequality} document, the values of noise
reduction factor $ R $ are practically independent of the mean
detected-photon numbers. On the other hand, quantities related to
the other two non-classicality criteria depend strongly on the
mean detected-photon numbers. Whereas the Schwarz inequality is
more suitable for detecting non-classicality for small mean
detected-photon numbers, the inequality based on higher-order
moments is preferred for larger mean detected-photon numbers.
In fact, this criterion is more sensitive to noise with
respect to the other two criteria because of the presence of
higher-order moments. As a consequence, when the mean numbers of
photons are very low, a lot of acquisitions is required for
successful application of this criterion.

\section{Nonclassical characteristics of the reconstructed photon fields}

The generated TWB states are highly nonclassical as they are
composed of photon pairs. The amount of their non-classicality
decreases during their propagation towards the detectors as some
of photons lose their twins. However, by far the largest loss of
non-classicality occurs during the detection by hybrid
photodetectors as their actual overall detection efficiencies lie
around 17$\%$, as confirmed by the minimum value achieved by $R$.
Despite this and in accordance with the results of the previous
Section, even the detected photons exhibit strong pairwise
correlations that guarantee nonclassical behavior of the
detected-photon fields. Nevertheless, the amount of
non-classicality found in the detected-photon fields is
considerably lower compared to that of the original TWB containing
photon pairs.\\
\begin{figure}   % fig. 7
\begin{center}
$(a)$\\
 \includegraphics[width=7.5cm]{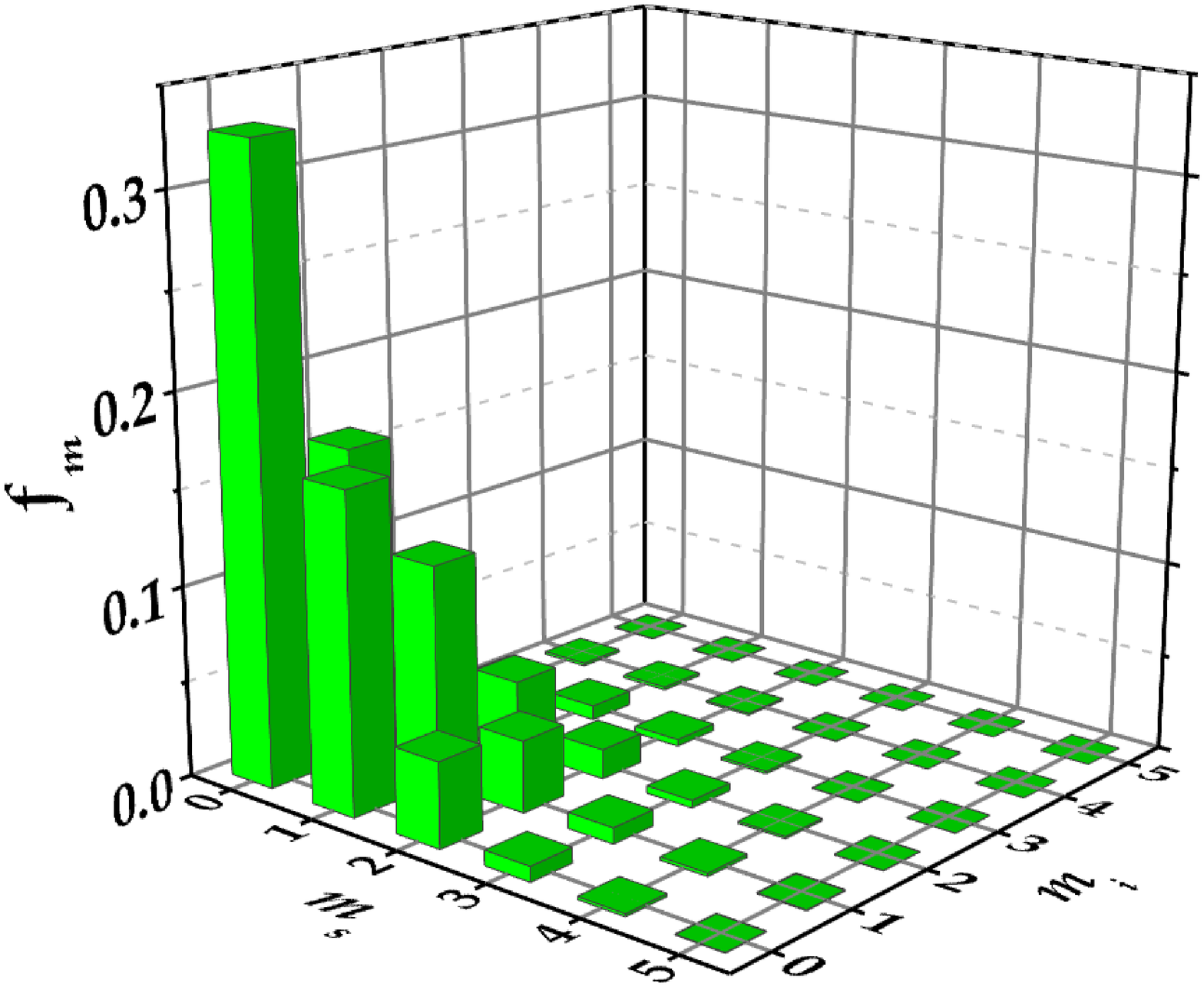}\\
 $(b)$\\
 \includegraphics[width=7.5cm]{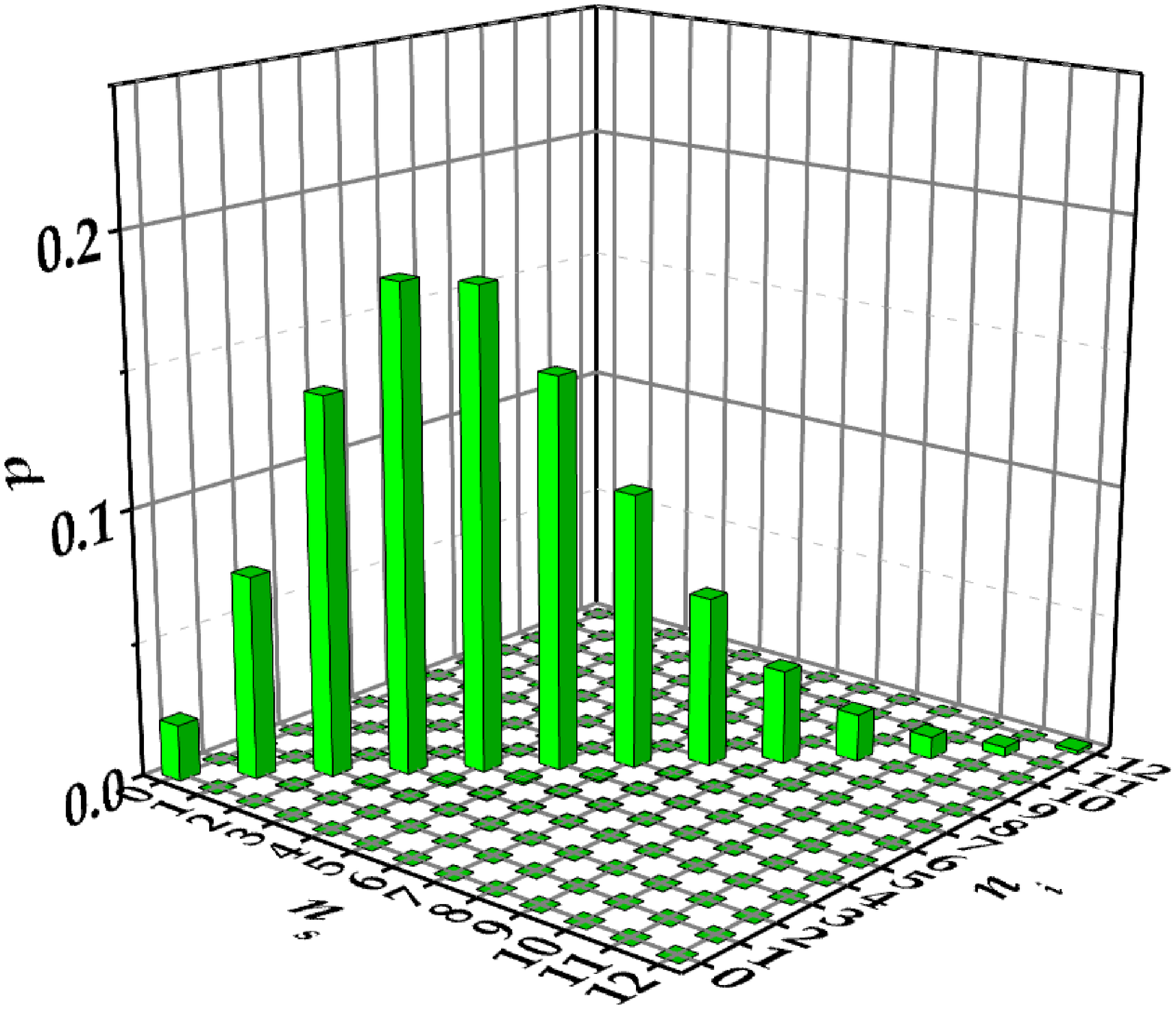}
 \end{center}
  \vspace{-0.6cm}
 \caption{Color online. $(a)$, experimental joint signal-idler detected
 photon-number distribution $f_m(m_s,m_i)$ and $(b)$, reconstructed joint signal-idler
 photon-number distribution $p(n_s,n_i)$ for the pump power
 49.2~$\mu$W.}
\label{fig7}
\end{figure}
For this reason, it is important to reconstruct the original TWB
in terms of photon numbers starting from the experimental
detected-photon distributions $f_m(m_s,m_i)$ in order to reveal
the quantum nature of state emitted in the nonlinear process. The
reconstructed joint signal-idler photon-number distributions $
p(n_s,n_i) $ can be obtained either by applying the
maximum-likelihood approach
\cite{Haderka2005a,PerinaJr2012,Perina2006,PerinaJr2012a} or by
fitting the experimental detected-photon distributions using a
special analytical form of the photon-number distribution $
p(n_s,n_i) $ \cite{Perina2005,Perina2006}. The second approach is
more convenient as it allows us to determine also quantum
detection efficiencies $\eta_s$ and $\eta_i$ of the signal and
idler beams, respectively \cite{PerinaJr2012a}.
The method only assumes
that the detected non-ideal TWB can be decomposed into three
statistically independent parts, namely the paired part, the
signal noise part and the idler noise part, which are all
described by multimode thermal fields. According to this model,
the joint signal-idler photon-number distribution $ p(n_s,n_i) $
\cite{Perina1991} can be written as
\begin{eqnarray}  % 6
 p(n_s,n_i) &=& \sum_{n=0}^{{\rm min}[n_s,n_i]} p_{\rm MR}(n_s-n;\mu_s,b_s)
  \nonumber \\
 & & \mbox{} \times
  p_{\rm MR}(n_i-n;\mu_i,b_i) p_{\rm MR}(n;\mu_p,b_p),
\label{6}
\end{eqnarray}
in which the Mandel-Rice distributions are written as $ p_{\rm
MR}(n;\mu,b) = \Gamma(n+\mu) / [n!\, \Gamma(\mu)]
b^n/(1+b)^{n+\mu} $ and $ \Gamma $ denotes the $ \Gamma
$-function. In Eq.~(\ref{6}), mean photon (photon-pair) numbers
per mode $ b_k $ and numbers $ \mu_k $ of independent modes for
the paired part ($ k=p $), noise signal part ($ k=s $) and noise
idler part ($ k=i $) as suitable characteristics of the analyzed
TWBs have been introduced. As the Mandel-Rice distributions in
Eq.~(\ref{6}) are defined for arbitrary nonnegative real numbers $
\mu $ of modes, the same applies also to the distribution $
p(n_s,n_i) $ in Eq.~(\ref{6}). This allows to consider a broader
class of analytic distributions when fitting the experimental
data. We note that the formula (\ref{eq:multithermal}) has been derived
for an integer number $ \mu $ of modes, but its generalization to real
nonnegative $ \mu $ is straightforward \cite{Perina1991}.\\
\begin{figure}   % fig. 8
\begin{center}
$(a)$\\
\includegraphics[width=7.5cm]{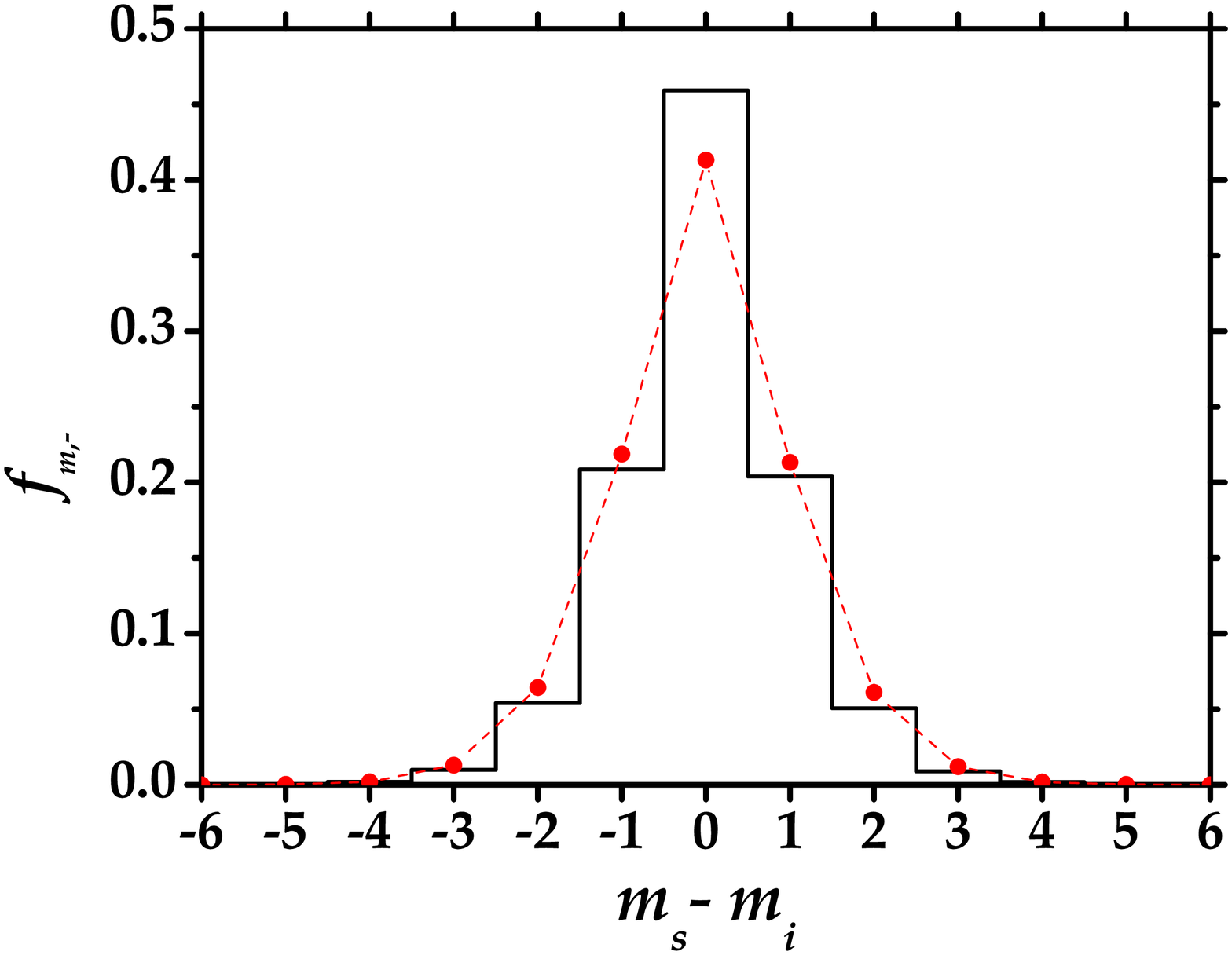}\\
$(b)$\\
\includegraphics[width=7.5cm]{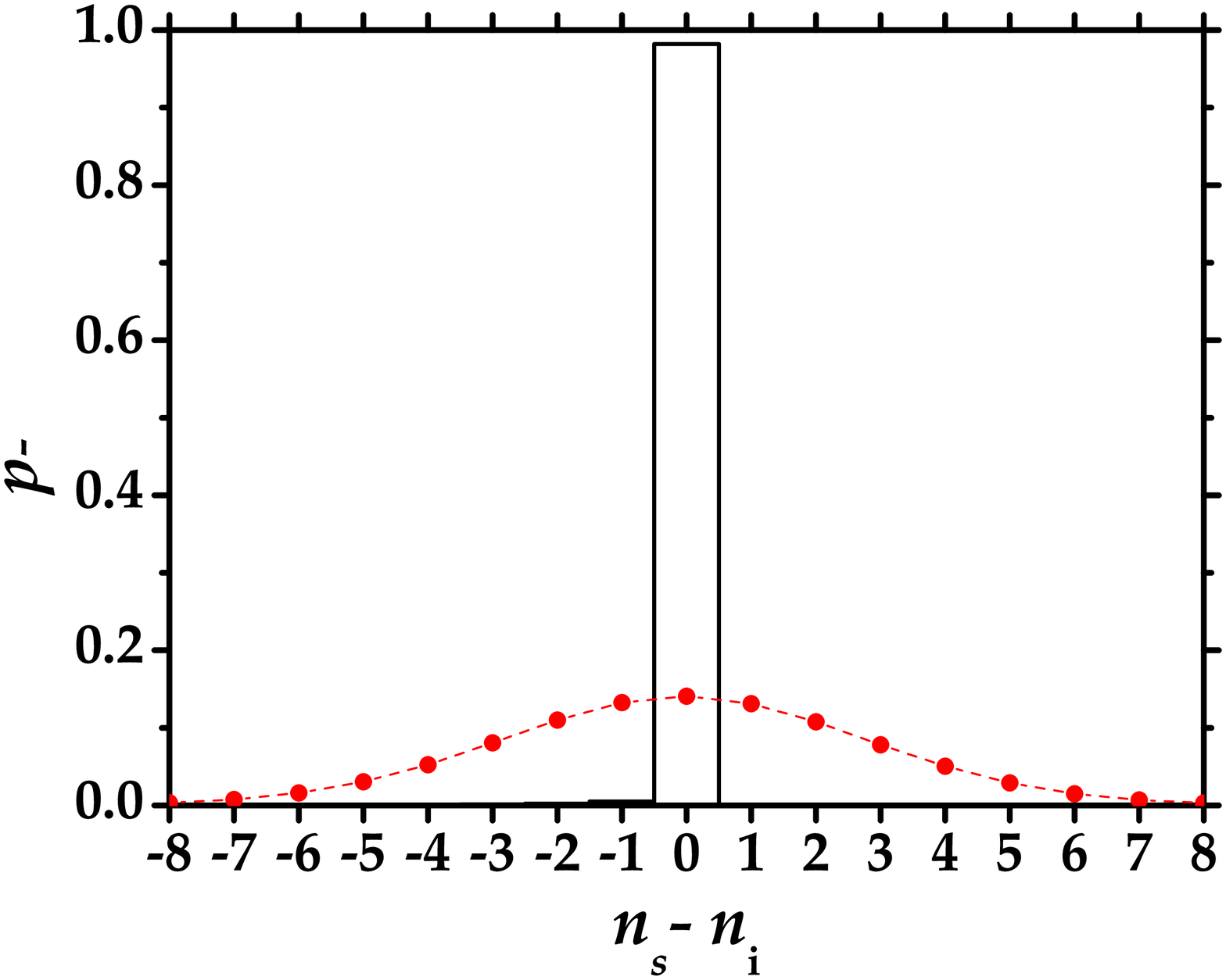}
 \end{center}
  \vspace{-0.6cm}
 \caption{Color online. $(a)$, detected photon-number distribution $ f_{m,-}(m_s-m_i) $ (bars) and $(b)$, photon-number
 distribution $ p_{-}(n_s-n_i) $ (bars) of the difference between signal and idler
 detected-photon and photon numbers, respectively,
 for the data shown in Fig.~\ref{fig7}. In the two panels we also show the distributions obtained by the
 combination of two independent classical fields with Poissonian statistics (dashed line + symbols).}
\label{fig8}
\end{figure}
The photon-number distribution $ p(n_s,n_i) $ is related to the
theoretical detected-photon distribution $ f_{m,th}(m_s,m_i) $ by
quantum detection efficiencies $ \eta_s $ and $ \eta_i $
\cite{PerinaJr2012}. Since detection by hybrid photodetectors is
characterized by the Bernoulli distribution, we can express this
relation as
\begin{eqnarray}   % 7
  f_{m,th}(m_s,m_i) = \sum_{n_s,n_i=0}^{\infty} B_s(m_s,n_s)
   B_i(m_i,n_i) p(n_s,n_i)
\label{7}
\end{eqnarray}
using the Bernoulli coefficients $ B_k(m_k,n_k) $,
\begin{equation}  % 8
 B_k(m_k,n_k) = \begin{pmatrix} n_k \cr m_k \end{pmatrix}
  \eta_k^{m_k} (1-\eta_k)^{n_k-m_k} .
\label{8}
\end{equation}
A fitting procedure that minimizes the declination between the
experimental histogram $ f_m(m_s,m_i) $ and theoretical
detected-photon distribution $ f_{m,th}(m_s,m_i) $ under the
assumption of equality of the first and second experimental and
theoretical detected photon-number moments (for details, see
\cite{PerinaJr2013}) allows us to determine both quantum detection
efficiencies $ \eta_k $, $ k=s,i $, and parameters $ b_k $ and $
\mu_k $, $ k=p,s,i $, of the analyzed TWB. To give a typical
example, we consider the experimental data obtained for pump mean
power 49.2~$ \mu$W and iris sizes' area 46~mm$^2$ (see the
marginal distribution plotted as black dots in
Fig.~\ref{statistics}). The fitting procedure assigned
the following parameters to the experimental distribution $ f_m $:
$ \eta_s = 0.147 $, $ \eta_i = 0.150 $, $ \mu_p = 31 $, $ b_p =
0.13 $, $ \mu_s = 1.2 \times 10^{-3} $, $ b_s = 24 $, $ \mu_i =
5.5 \times 10^{-3} $, and $ b_i = 13 $. 
First of all, we note that the values of quantum efficiencies obtained by 
the reconstruction method are comparable with the value obtained from the noise reduction
factor for the same set of data (see points at 46~mm$^2$ in Fig.~\ref{NRF}(b)) \cite{bridaOE}.
Second, we remark that the paired part of TWB
representing more than 98\% of the entire field is 
described by a multi-thermal field with 31 independent modes. We
note that the mean number of photons in paired fields equals 8, whereas the means
of noisy signal and idler photon numbers lay below 0.1. On the
other hand, the noise signal and idler parts have numbers $ \mu $
of modes much less than one which means that their probability
densities have appreciated values only very close to the zero
photon number. This is a consequence of very low noise signal and
idler intensities observed in the experiment. We attribute the
found numbers $ \mu $ of modes much less than one to distortions
of electronic signals inside the detection chains
including HPDs.\\
Finally, we point out that whereas the joint signal-idler experimental
detected-photon histogram $ f_m $ provided covariance equal to
0.16, covariance of photon numbers in the reconstructed
photon-number distribution $ p $ is equal to 0.85. The
reconstruction also decreased the value of noise reduction factor
$ R $ to 0.2. This dramatic increase of correlations between the
signal and idler fields in a TWB after the reconstruction also
changes the shape of the corresponding joint signal-idler
(detected) photon-number distributions (see Fig.~\ref{fig7}).
In fact, the presence of nonzero off-diagonal elements
in the detected photon-number distribution in Fig.~\ref{fig7}$(a)$
makes its nonclassical character less evident compared to the
reconstructed photon-number distribution $ p(n_s,n_i) $ plotted in
Fig.~\ref{fig7}$(b)$ and clearly showing the prevailing pairwise
character of the TWB (the off-diagonal elements attain values
lower than 1\% of those of diagonal elements). Also, the sum of
diagonal elements gives 98.2\% of the entire joint signal-idler
photon-number distribution. This is in accord with the relative weights
of paired, noise signal and noise idler parts of the TWB expressed
in mean pair/photon numbers.
\begin{figure}   % fig. 9
\begin{center}
$(a)$\\
 \includegraphics[width=7.5cm]{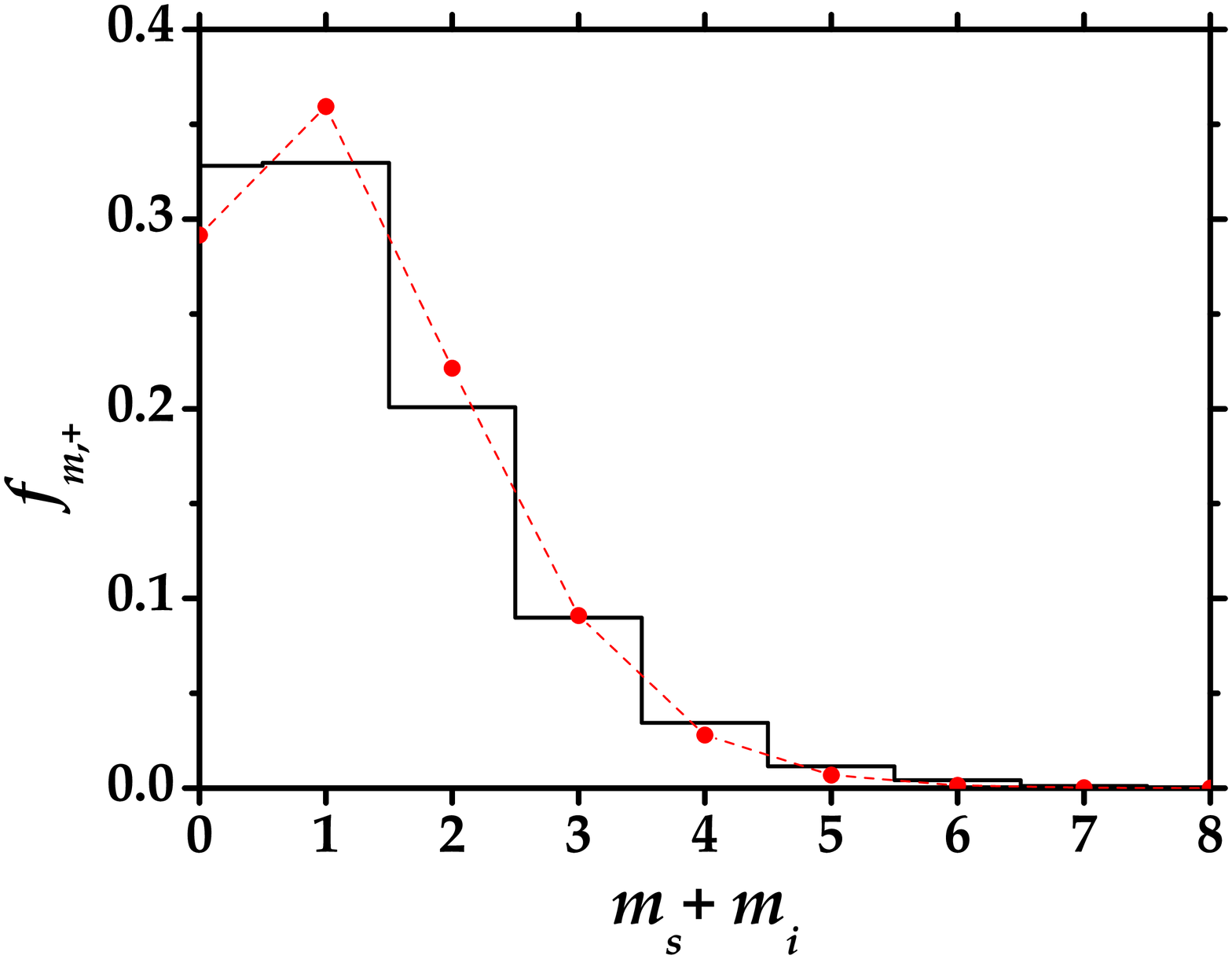}\\
 $(b)$\\
\includegraphics[width=7.5cm]{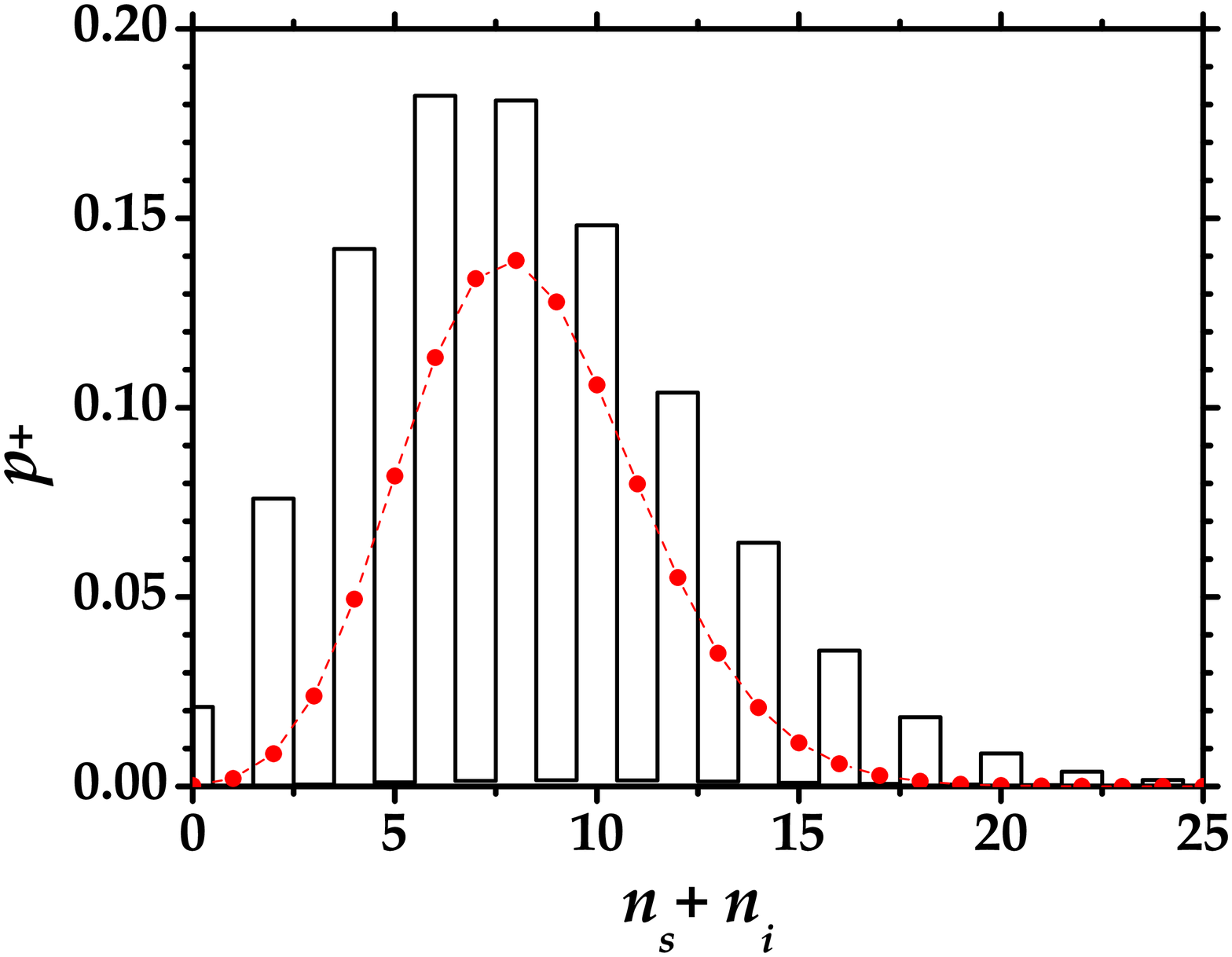}
 \end{center}
  \vspace{-0.6cm}
 \caption{Color online. $(a)$, detected photon-number distribution $ f_{m,+}(m_s+m_i) $ (bars) and $(b)$, photon-number
 distribution $ p_{+}(n_s+n_i) $ (bars) of the sum of signal and idler detected-photon and photon numbers,
 respectively, for the data shown in Fig.~\ref{fig7}. In the two panels we also show the distributions obtained by the
 combination of two independent classical fields with Poissonian statistics (dashed line + symbols).} \label{fig9}
\end{figure}
A substantial difference in the nonclassical behavior of
detected-photon-number and photon-number distributions can be
observed in the corresponding distributions of the sum and
difference of the signal and idler detected-photon and photon
numbers, respectively. The resulting distributions are compared
with those obtained by the combination of two independent
classical fields with Poissonian statistics. This comparison
applied to the experimental detected-photon distribution reveals
only weak signatures of non-classicality in the distributions $
f_{m,+}(m_s+m_i) $ and $ f_{m,-}(m_s-m_i) $ of the sum $ m_s+m_i $
and difference $ m_s-m_i $ of the signal and idler detected-photon
numbers defined as:
\begin{eqnarray} % 9
 f_{m,+}(m) &=& \sum_{m_s,m_i=0}^{\infty} \delta_{m,m_s+m_i} f_m(m_s,m_i)
  , \nonumber \\
 f_{m,-}(m) &=& \sum_{m_s,m_i=0}^{\infty} \delta_{m,m_s-m_i}
  f_m(m_s,m_i) ,
\label{9}
\end{eqnarray}
where $ \delta $ denotes the Kronecker symbol. As shown in
Fig.~\ref{fig8}$(a)$, the experimental distribution $
f_{m,-}(m_s-m_i) $ of the difference is slightly narrower than the
reference distribution. On the other hand, a slightly broader
experimental distribution $ f_{m,+}(m_s+m_i) $ of the sum with
respect to the reference distribution is drawn in
Fig.~\ref{fig9}$(a)$. The reconstruction of joint photon-number
distribution clearly reveals non-classicality of TWBs, as
documented by the photon-number distributions $ p_{-}(n_s-n_i) $
and $ p_{+}(n_s+n_i) $ plotted in Figs.~\ref{fig8}$(b)$ and
\ref{fig9}$(b)$. The distribution $ p_{-}(n_s-n_i) $ of
photon-number difference plotted in Fig.~\ref{fig8}$(b)$
demonstrates the prevailing pairwise character of TWBs that is
also confirmed by a 'teeth-like' character of the photon-number
distribution $ p_{+}(n_s+n_i) $ of photon-number sum depicted in
Fig.~\ref{fig9}$(b)$.
\begin{figure}   % fig. 9
\begin{center}
$(a)$\\
\includegraphics[width=7.5cm]{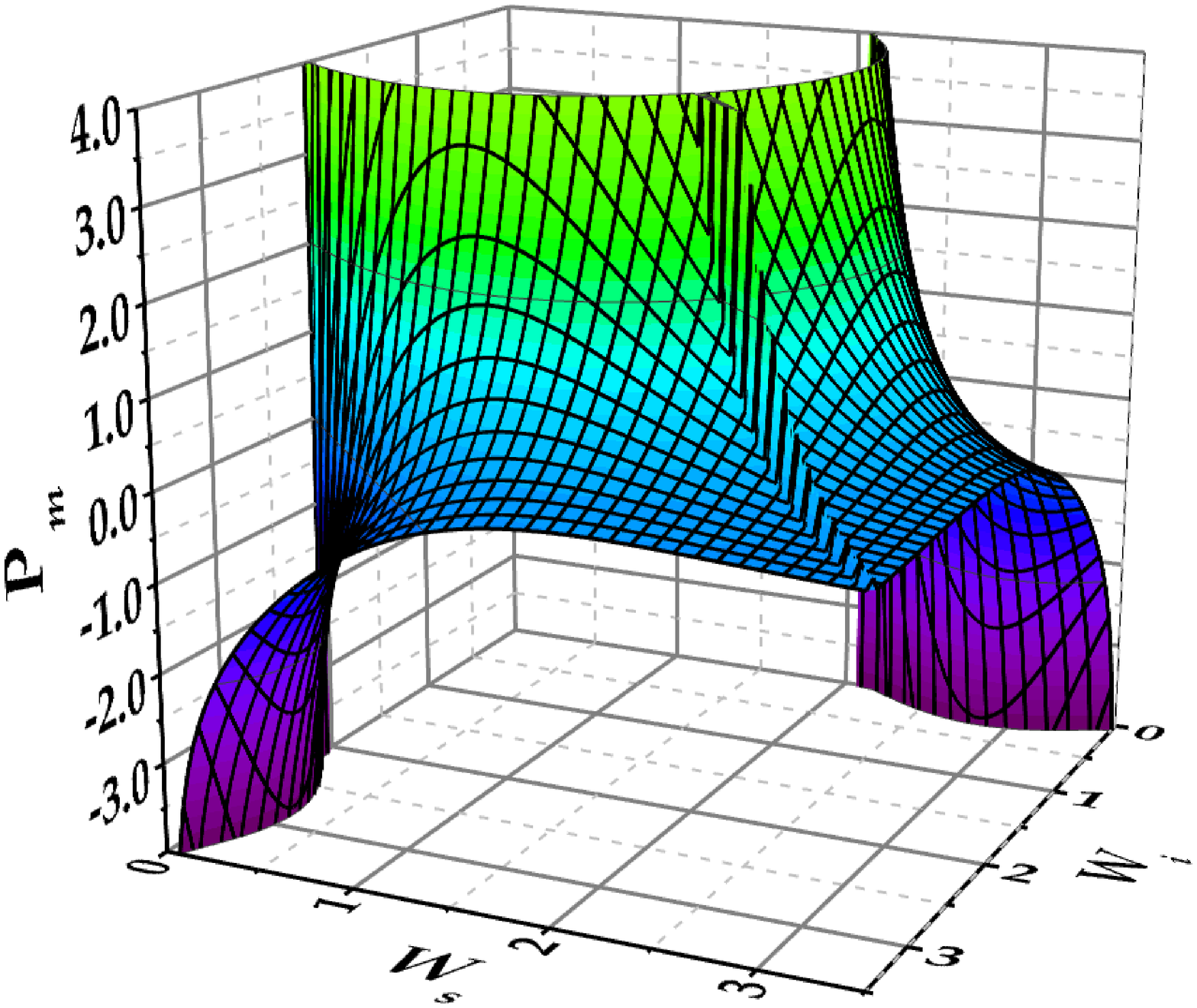}\\
 \vspace{-0.3cm}
$(b)$\\
\includegraphics[width=7.5cm]{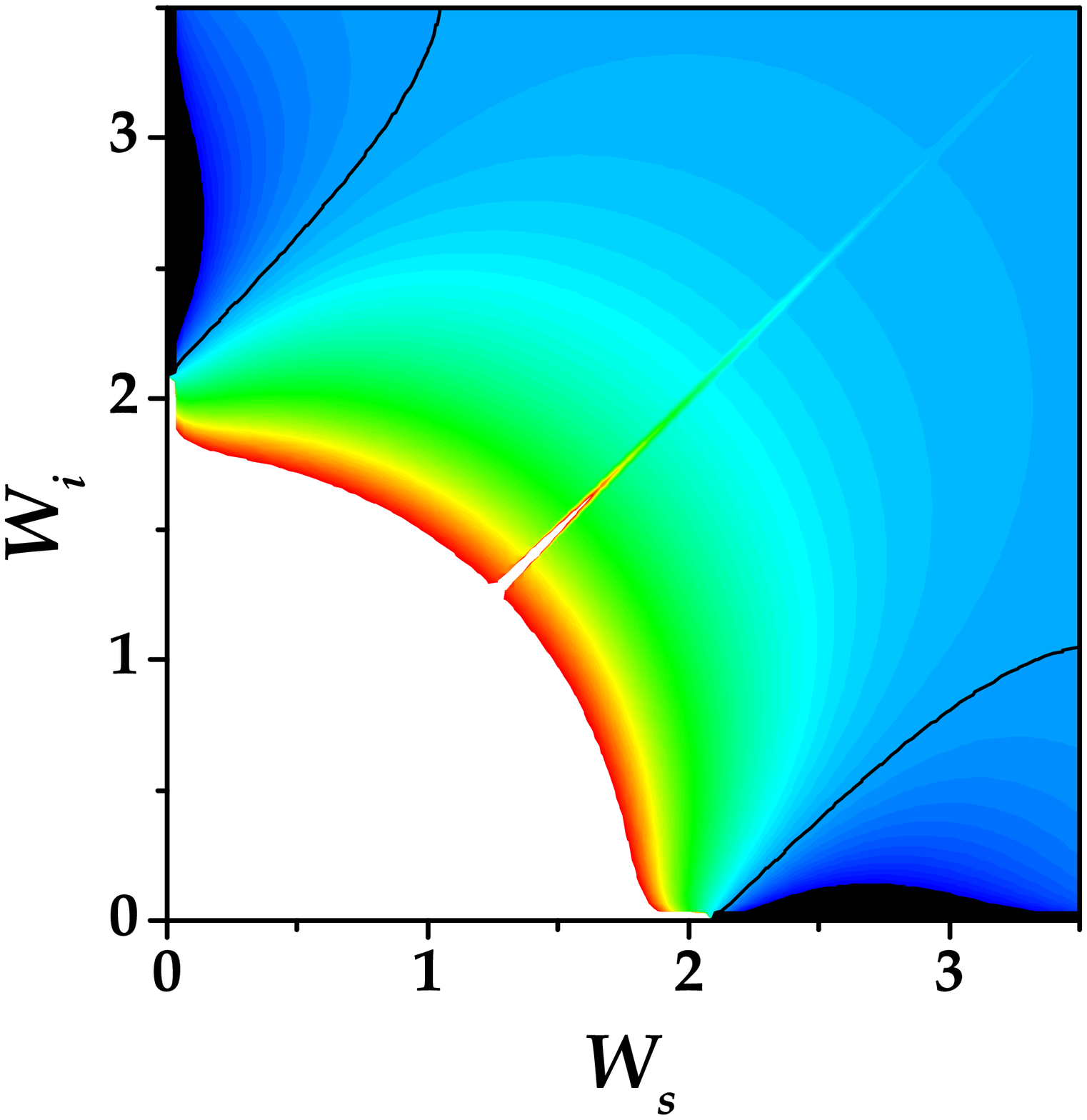}\\
 \vspace{-0.3cm}
$(c)$\\
\includegraphics[width=7.5cm]{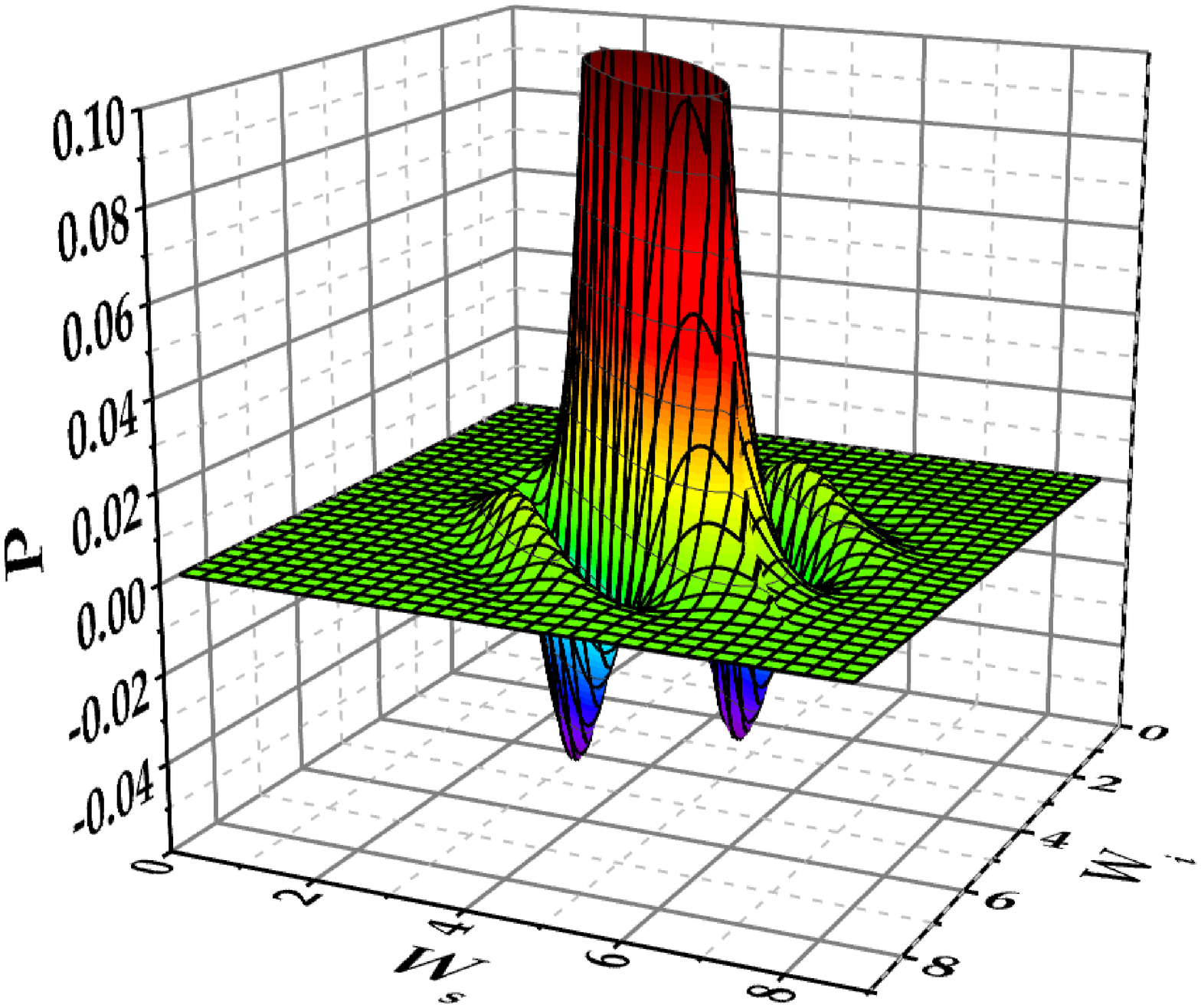}\\
 \vspace{-0.3cm}
$(d)$\\
\includegraphics[width=7.5cm]{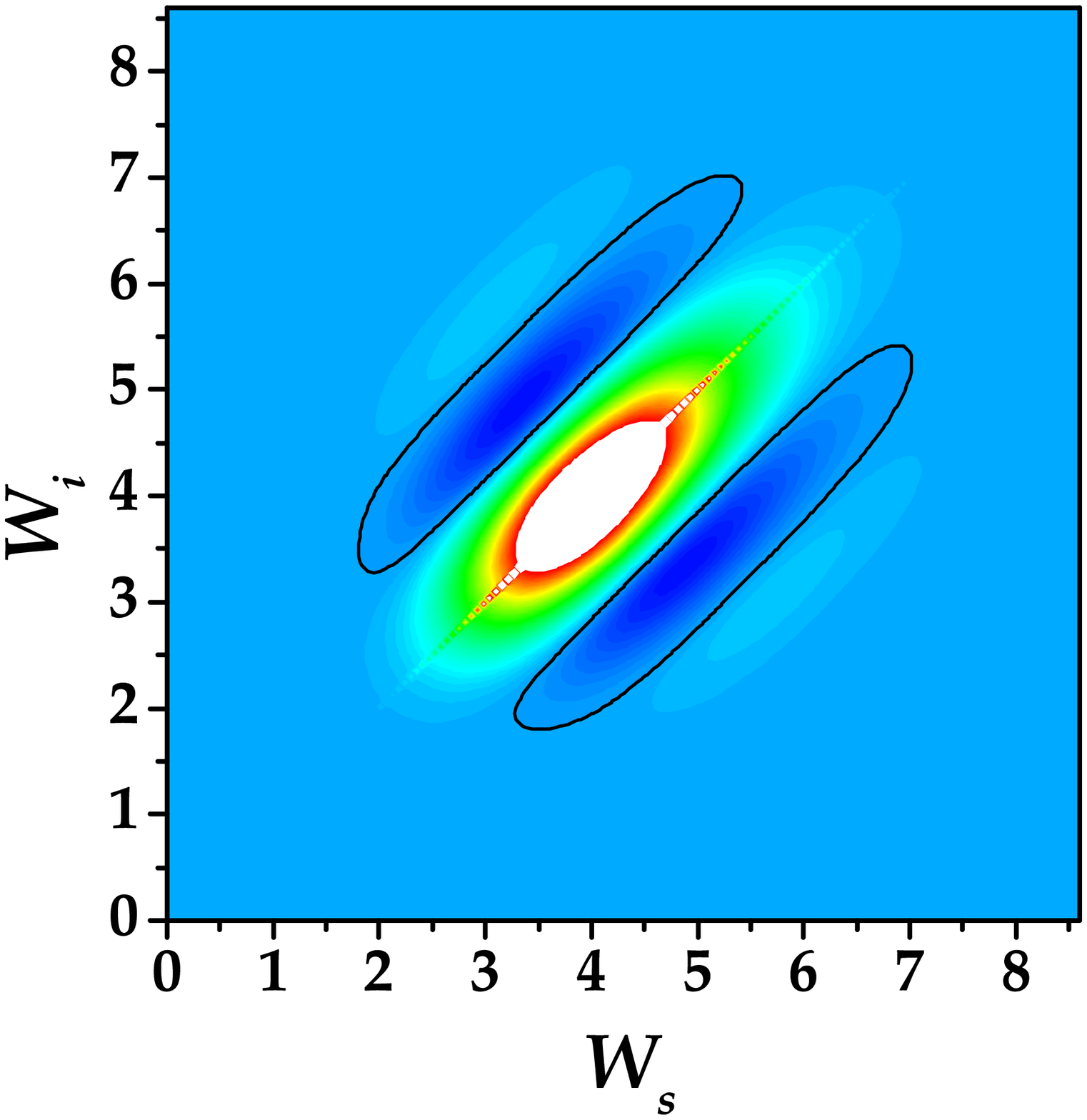}
 \end{center}
 \vspace{-0.8cm}
 \caption{Color online. $(a)$, quasi-distribution of
 `detected-photon intensities' $ P_{m}(W_{m,s},W_{m,i}) $ and its topo graph, $(b).$ $
 (c)$, quasi-distribution of photon integrated intensities  $ P(W_{s},W_{i}) $ and its topo graph,
 $(d)$. Topo graphs in $(b)$ and $(d)$ have the same scales as in $(a)$ and $(c)$, respectively.
 In $(b)$ and $(d)$ black contours mark the zero level.}
\label{fig10}
\end{figure}
\\
An ultimate criterion for discriminating quantum and classical
multimode fields is related to the properties of
quasi-distribution $ P $ of integrated intensities, i.e.
electric-field intensities integrated over the detection interval,
related to normal ordering of field operators (for more details,
see, e.g., \cite{Perina1991,mandel,Perina2011}). The reason is
that integrated intensities describe the fields before detection
that may conceal nonclassical features of these fields. The
relation between integrated intensities and detected photons is
provided by Mandel's detection formula \cite{mandel}. This formula
can be inverted \cite{Perina1991} and then used for the
determination of quasi-distributions of integrated intensities
from the photon-number distributions obtained from experimental
data. According to quantum theory of radiation
\cite{Glauber1963,Perina1991} if the quasi-distribution $ P $
attains negative values or is even singular, the field is
nonclassical. The quasi-distribution $ P(W_s,W_i) $ of signal ($
W_s $) and idler ($ W_i $) integrated intensities can be written
in the form of two-fold convolution, which is a consequence of
Eq.~(\ref{6}) for the photon-number distribution $ p(n_s,n_i) $
\cite{PerinaJr2013}:
\begin{eqnarray}  % 10
 P(W_s,W_i) \hspace{-2mm} &=& \hspace{-2mm} \int_{0}^{\infty}
  dW'_s \int_{0}^{\infty} dW'_i
  P_{p}(W_s-W'_s,W_i-W'_i) \nonumber \\
 & & \mbox{} \times
  P_{s}(W'_s) P_{i}(W'_i).
\label{10}
\end{eqnarray}
Quasi-distributions $ P_k $ of integrated intensities introduced
in Eq.~(\ref{10}) describe the paired ($ k=p $), signal noise ($
k=s $) and idler noise ($ k=i $) parts of the TWB. More details
can be found in \cite{Perina2005,PerinaJr2013}.\\
As we have demonstrated, many non-classicality criteria indicate
quantum behavior of even experimental distributions written in
terms of detected photons. Following the genuine definition of
non-classicality, we can define a quasi-distribution $ P_m $ of
`detected-photon intensities' following the approach developed for
photons and assuming perfect quantum detection efficiencies ($
\eta_s = \eta_i = 1 $) \cite{Perina2007}. Of course, the obtained
quasi-distribution $ P_m $ characterizes a fictitious
`detected-photon' boson field, as it contains only those photons
that are captured by the detectors. As in the case of
quasi-distribution of integrated intensities, the existence of
negative regions in the quasi-distribution $ P_m $ for detected
photons confirms the nonclassical character of the state. The
quasi-distribution $ P_m(W_{m,s},W_{m,i}) $ of `detected-photon
intensities' determined from the analyzed experimental
distribution $f_m$ is shown in Fig.~\ref{fig10}. In order to see a
detained behavior of this quasi-distribution and in particular to
investigate in which regions it attains values close to zero, we
plot only a part of the function in Fig.~\ref{fig10}$(a)$ and remark
that the maximum of the peak in the origin reaches the value
7$\times 10^5$. The smallest negative values, equal to -0.2, are
found close to the $ W_{m,s} $ and $ W_{m,i} $ axes. The highly
prevailing positive part of quasi-distribution $ P_m $ indicates
that the measured state is close to a classical one. However, the
presence of a negative part (even small) shows that the low
detection efficiency has preserved the pairwise character of TWB.
The comparison of the quasi-distribution $ P_m $ of
`detected-photon intensities' with the genuine quasi-distribution
$ P $ of photon intensities [see Fig.~\ref{fig10}$(c)$] reveals much
stronger non-classicality in the case of photons. We note that the
peak value of $ P $ in Fig.~\ref{fig10}$(c)$ equals 0.99 which is
considerably lower than the peak value of quasi-distribution $ P_m $ shown in
Fig.~\ref{fig10}$(a)$. Nevertheless, both quasi-distributions attain
negative values and so both describe a nonclassical field. The
contour plots of both quasi-distributions depicted in
Figs.~\ref{fig10}$(b)$ and 10$(d)$ reveal that negative values of
these distributions are localized in parallel strips whose
orientation originates in the pairwise character of TWBs.

\section{Conclusions}

Using spontaneous parametric down-conversion in the linear gain
regime, we generated multimode twin-beam states in the mesoscopic
photon-number regime. We studied nonclassical properties of the
twin beams by applying three different non-classicality criteria
written in terms of detected photons. Whereas the noise reduction
factor $ R $ is a suitable indicator of non-classicality
independent of the twin-beam intensity, the Schwarz inequality
is useful for weak twin beams and the criterion derived from
higher-order detected-photon-number moments finds its application
for intense twin beams. To compare these criteria with the genuine
definition of non-classicality we also determined
quasi-distributions of detected-photon and photon integrated
intensities for normally ordered field operators. Despite the low
detection efficiency (around $17\%$) negative values of these
quasi-distributions found in typical strips were observed both for
photons and detected photons, confirming non-classicality of the
generated twin beams. The set of criteria we presented can thus be
considered as a robust tool for quantifying non-classicality of
multimode twin beams used in many schemes, including that for
conditional generation of nonclassical and non-Gaussian states.

\section{Acknowledgements}
The research leading to these results has been supported by MIUR
(FIRB ÒLiCHISÓ - RBFR10YQ3H). Support by projects P205/12/0382 of
GA \v{C}R, Operational Program Research and Development for
Innovations - European Regional Development Fund project
CZ.1.05/2.1.00/03.0058 and Operational Program Education for
Competitiveness - European Social Fund project
CZ.1.07/2.3.00/20.0058 of M\v{S}MT \v{C}R are acknowledged.

\end{document}